# Simulation of flow-induced vibration of a cylinder in expansion tube

Sai Peng, Qiyu Deng, Lin Zhou, Tao Huang and Peng Yu*

[1]Department of Mechanics and Aerospace Engineering, Southern University of Science and Technology, Shenzhen, 518055, China

**Abstract**

In this study, a series of simulations are conducted to investigate the motion of a small cylinder in an expansion tube, focusing on two-dimensional dynamics. These simulations are performed on the FLUENT platform employing the Overset function. The collision between the cylinder and the tube wall is modeled as a positive rigid body collision without losing energy. Two key parameters, the dimensionless gravity ($Mg^*$) and the Reynolds number ($Re$), were explored within the ranges of 4.9-79.4 and 1-300, respectively. Two types of inflow are considered: the invariable inflow or superimposed a sinusoidal inflow of periodical fluctuation. For invariable inflow, three motion modes (drainage mode, balance mode and vibration mode) are found in the phase diagrams of ($Mg^*$, $Re$). For an invariable inflow, the high-amplitude vibrations of a cylinder is proved to be widespread in the Reynolds number range from 5 to 40. Additionally, the scenario involving $Re$ =300, $Mg^*$ =39.25 with sinusoidal periodic fluctuation incoming flow demonstrated intense vibrations of the cylinder. The behavior of the two-dimensional cylinder can be approximated as a simplified version of a three-dimensional sphere when the spherical motion is disregarded. Our research may help to understand this ancient flow problem.

## 1. Introduction

Flow induced vibration (FIV) is a classical fluid-structure coupling phenomenon that has drawn considerable attention due to its complexity and wide-ranging applications, such as structural fatigue failure (Xu et al., 2020), vibration noise (Hattori et al., 2017), and enhanced heat dissipation (Lu et al., 2010; Kumar et al., 2020). The FIV often simplifies the vibration system into a mass-spring system. Moreover, the rigid body system influenced solely by gravity can induce vibrations, as observed in systems like the single ball-pendulum system and small ball in the expansion tube, etc. The movement of a small ball in expansion tube is frequently used as an illustrative example in the explanation of the Bernoulli equation in the teaching of fluid mechanics. A bottle in a expansion tube presents a straightforward system; however, understanding the bottle's motion in the presence of wake vortex shedding, variable blockage rates, vibrations, and collisions can be challenging. In this study, a two-dimensional simulation of a small cylinder in an expansion tube is conducted, as illustrated in Fig. 1.

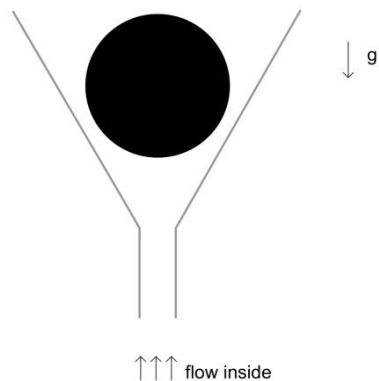

**Figure 1.** A small cylinder in an expansion tube.

The flow-induced vibration is a result of the interaction between the cylinder's motion and the shedding of separated vortices (Williamson & Govardhan, 2004). Understanding the movement of the cylinder is crucially linked to the wake vortex shedding. The Reynolds number at which wake vortex shedding initiates over a cylinder is influenced by various factors including vibration of a cylinder (Kou et al., 2017), rotation (Kang et al., 1999, Rao et al., 2015), and blocking rate (Sahin and Owens, 2004), etc. The critical Reynolds number for a cylinder's vortex shedding in a free domain is typically about 47 (Lashgari et al., 2012). However, vibration of the cylinder can significantly decrease this critical Reynolds number (Kou et al., 2017). Additionally, the cylinder's blockage rate also affects critical Reynolds number. Sahin and Owens (2004) extensively studied the influence of the blocking rate on critical Reynolds number. In our system, the critical Reynolds number of the cylinder is influenced not only by the vibrations but also the gradient of the blockage rate.

In this system, the collision play a significant role and cannot be overlooked. When simulating collisions, empirical equations of lubrication forces (Izard et al., 2014) or reflective boundary (Ardekani & Rangel, 2008) are commonly employed to ensure rapid separation of wall faces without causing non-physical penetration. Fluid-solid coupling often utilizes methods such as mesh reconstruction (Xiong et al., 2019), immersed boundary techniques (Izard et al., 2014), or Overset method (Chandar et al., 2018). While the mesh reconstruction method may not be suitable for handling topological changes or large motion problems. The immersed boundary method or Overset method may be used to better address collision issues. Compared with the immersed boundary method, the Overset method can maintain the mesh quality near the wall. The Overset method employs two series of grids, one series of background grids and the other series of component grids. The background grid does not move, and the component grid moves, interpolated to the background grid. In this study, the Overset method is employed. It is worth noting that particle-wall impacts hold significance in other areas of fluid mechanics such as the enhancement of heat transfer due to convection of fluid towards and away from a surface and for the development of improved multi-phase models that include wall effects.

Comprehensive reports on the vibration of the two-dimensional cylinder or three-dimensional sphere in the expansion nozzle are lacking. Theoretical and experimental research by Masmoudi et al. (1998) focused on final stage of sedimentation of a spherical particle moving along the axis of a conical vessel with a viscous incompressible fluid. They observed that the particle settling velocity varies like $d^{5/2}$, where $d$ represents the gap. The study highlighted a close agreement between this result from the lubrication theory and experiment. The creeping motion of a sphere along the axis of a closed axisymmetric container has been researched by Lecoq et al. (2007). They used the numerical technique to solve the Stokes equation using the classical Sampson expansion and experimental technique to get vertical displacement by using laser interferometry with an accuracy of $50nm$. Both studies were conducted at a very low Reynolds number ($Re$) and do not consider the incoming flow stream from the bottom.

In this work, we employ the Overset method to conduct a novel simulation of a cylinder in the expansion tube. The simulation investigates the cylinder's movements within a unique expansion tube, focusing on analyzing trajectories, vibration amplitudes, and different forms of movement. Mathematical formulations will be discussed to provide insight into the research approach.

## 2. Model equations and specific problem

### 2.1. 2.1 Model equations

In this study, the Arbitrary Lagrangian–Eulerian (ALE) scheme is applied to simulate the movement of the boundaries due to the motion of the cylinder. The hydrodynamics of incompressible fluids can be described by the following set of equations (Zhao, 2020):

$$\nabla \cdot (\mathbf{u} - \mathbf{u}_c) = 0, \tag{1}$$

$$\rho_f \frac{\partial \mathbf{u}}{\partial t} + \rho_f \left[ (\mathbf{u} - \mathbf{u}_c) \cdot \nabla \right] \mathbf{u} = -\nabla p + \mu \Delta \mathbf{u}. \tag{2}$$

Where $\mathbf{u}$ is the two-dimensional velocity vector. $t$ is the time. $\mathbf{u}_c$ is the velocity vector of the moving grid. $p$ is pressure. $\mu$ is the viscosity. $\rho_f$ is density of fluid. The Reynolds number is defined as $Re = \rho_f D U_\infty / \mu$, where $D$ is the cylinder's diameter. The gravity of fluid is not considered in this equation.

The generalized solid movement for two degrees of freedoms can be written as follows:

$$\mathbf{F}^*(t) = m \frac{d^2 \mathbf{x}}{dt^2} + (0, mg), \tag{3}$$

where $\mathbf{x} = (x, y)$, $x$, $y$ are the horizontal and transversal displacement, respectively, $\mathbf{F}^*(t) = (F_x^*(t), F_y^*(t))$, $F_x^*(t)$, $F_y^*(t)$ are the horizontal and transversal forces acting on the cylinder, respectively, $m$ represents the mass of the cylinder, expressed as:

$$m = \rho_s \pi \left(\frac{D}{2}\right)^2, \tag{4}$$

where $\rho_s$ is density of the cylinder.

The fluid force acting on the cylinder comprises two components: the differential pressure force and the viscous force, which can be expressed as follows:

$$\mathbf{F}^* = \oint (-p^* \mathbf{I} + \mathbf{\tau}) \cdot d\mathbf{S}, \tag{5}$$

where $p^*$ and $\tau$ are the pressure and shear stress acting at cylinder, respectively. $\mathbf{I}$ is second-order two-dimensional unit tensor. $\mathbf{S}$ is the surface normal vector of the cylinder. The pressure $p^*$ is different to $p$ for the existence of fluid's gravity. In fact, the following conversion relationship exists between these two variables:

$$p^* = p - \rho_f g y. \tag{6}$$

The force of the fluid on the cylinder also can be written as follow:

$$\mathbf{F}^* = \oint [-p\mathbf{I} + \tau] \cdot d\mathbf{S} + \left(0,\ \rho_f g \pi D^2/4\right). \tag{7}$$

The force of the fluid on the cylinder if not consider gravity of the fluid can be written as follow:

$$\mathbf{F} = \oint [-p\mathbf{I} + \tau] \cdot d\mathbf{S}. \tag{8}$$

Introducing the fore coefficient $\mathbf{C}(Re, \mathbf{u}_c/\bar{U}) = (C_x(Re, \mathbf{u}_c/\bar{U}), C_y(Re, \mathbf{u}_c/\bar{U}))$, can be written as follow:

$$\mathbf{C}(Re, \mathbf{u}_c/\bar{U}) = 8\mathbf{F}/\rho_f \pi D^2. \tag{9}$$

The fore coefficient is a function of Reynolds number and the moving speed of the cylinder.

Combined Eqns (3), (4), (7), (8) and (9), we have

$$\frac{1}{\bar{U}^2/D}\left(1 - \frac{\rho_f}{\rho_s}\right)(0,\ g) - \frac{2\rho_f \mathbf{C}(Re, \mathbf{u}_c/\bar{U})}{\rho_s \pi} = \frac{D}{\bar{U}^2} \cdot \frac{d^2 \mathbf{x}}{dt^2}. \tag{10}$$

Let

$$g^* = \frac{g}{\bar{U}^2/D}, \tag{11-1}$$

$$M = \frac{\rho_s}{\rho_f} - 1, \tag{11-2}$$

$$\mathbf{x}' = \mathbf{x}/D, \tag{11-3}$$

$$t' = \frac{t}{D/\bar{U}}. \tag{11-4}$$

We have

$$(0,\ Mg^*) - (M+1)\frac{d^2 \mathbf{x}'}{dt'^2} = \frac{2\mathbf{C}(Re,\ \mathbf{u}_c/\bar{U})}{\pi}. \tag{12}$$

If the cylinder finally still somewhere, the Eqn. (12) can be simplified to

$$\mathbf{C}(Re) = \frac{\pi}{2}(0, Mg^*). \tag{13}$$

This indicates that relationship between the force coefficient and dimensionless gravity is balanced. Taking into consideration the Eqn. (12), in the scenario where the cylinder is in constant vibration, two dimensionless parameters $Mg^*$ and $\mathbf{u}_c/\overline{U}$ play a significant role. In fact, the movement speed of the solid body is influenced by both the $Re$ and $Mg^*$.

### 2.2. Simulation method description

In order to couple the structural displacement, a loose fluid structure coupling is adopted to simulate the non-linear FSI problem. This method involves obtaining the flow field based on instantaneous geometrical configuration and fluid governing equations. Subsequently, the hydrodynamic load is shared to the structural motion Eqn. (3). The structural displacement is then computed within the same physical time step based on an initial value problem of first-order ordinary differential equations through the state-space method. The pressure velocity coupling algorithm is adopted to solve the mass and momentum equations, yielding velocity and pressure results. Iterations proceed within each time step until the residual converges, determining the load exerted by the flow field on the cylinder. The velocity of the circular cylinder is updated to solve the displacement equations of the cylinders by a fourth order Runge–Kutta method in the same step, allowing for the determination of the circular cylinder's position and velocity at that instant.

The numerical solution to the governing equations is achieved using a control volume method, where partial differential equations are converted to a set of discrete algebraic equations with conservative property. A second-order upwind scheme is applied to discretize the convective term in the momentum equations, while a second-order central scheme is adopted to discretize the diffusive terms in momentum equations. Subsequently, the grid is updated and prepared for the subsequent time step. This process combines the computation of the flow field with the grid update, ensuring accurate and efficient simulation..

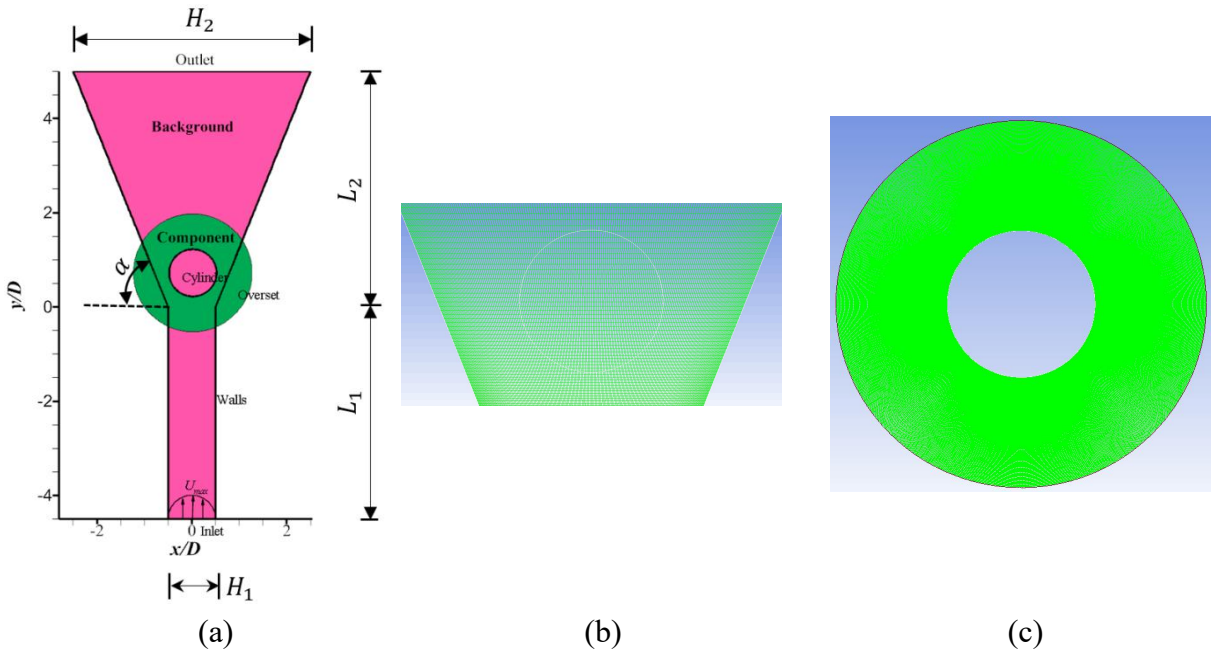

(a)                                    (b)                                   (c)

**Figure 2.** The grid of component and background.

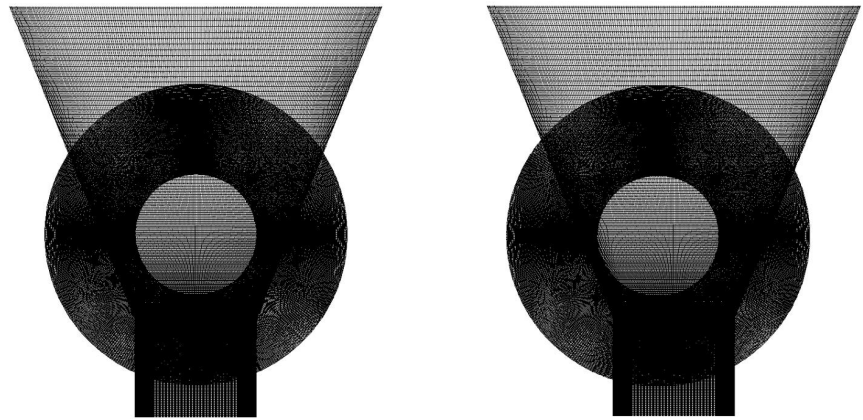

**Figure 3.** The grid at two moments.

Moving grids are managed using the Overset method.,which involves two sets of grids depicted in Fig. 2. The first set is background mesh , highlighted in the read region in Fig. 2(a) and Fig. 2(b), and remains unchanged throughout the calculation. The second set is the component mesh surrounding the cylinder, shown in green region in Fig. 2(a) and Fig. 2(c). This mesh set also remains undeformed and only moves with the cylinder. The moving component grid is then interpolated with the background grid. We select the grid at two moments as shown in Fig. 3. In the event of a collision between the cylinder and outer wall . We consider a scenario where the dry coefficient of restitution is zero. In cases of frictionless collision, the tangential force is zero and the same normal force applies to cylinder in opposite directions, as shown in Fig.

4. Speed of circular motion after reflection can be written as follow:

$$u_{ox} = \frac{2u_{iy}(x_1-x_0)(y_1-y_0)+u_{ix}(x_1-x_0)^2 - u_{ix}(y_1-y_0)^2}{(x_1-x_0)^2+(y_1-y_0)^2}$$
$$u_{oy} = \frac{2u_{ix}(x_1-x_0)(y_1-y_0)-u_{iy}(x_1-x_0)^2 + u_{iy}(y_1-y_0)^2}{(x_1-x_0)^2+(y_1-y_0)^2}$$
(14)

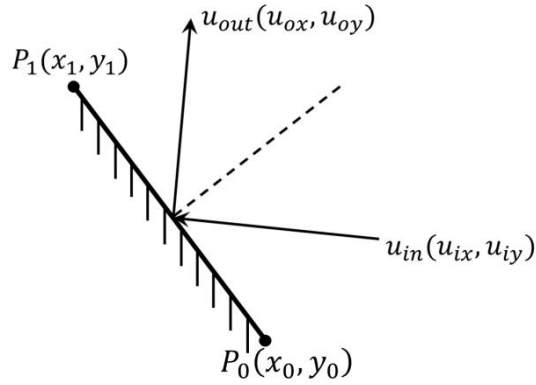

**Figure 4.** Reflection model.

### 2.3. Problem description

The calculated spatial domain and geometric dimensions are depicted in Fig. 2(a). At the initial moment, the cylinder is placed at the center point coordinates (0, 0.5$D$). The cylinder and the outer wall surface are set as a non-slip wall boundary. The pressure of outlet boundary is set as zero. The width of inlet boundary is set as $H_2 = 4D$. The outlet length is set as $L_2 = 5D$ in y-direction. The angle $\alpha$ satisfies $\tan(\alpha) = 3.3333$, that is, $\alpha$ equals to 73.3°. The width of inlet boundary is set as $H_1 = D$. The inlet length is set as $L_1 = 4.5D$. The inlet velocity is set as parabolic distribution:

$$u_y(x) = 1.5\bar{U}\left(1 - \frac{y^2}{H_1^2}\right).$$
(15)

Or the inlet velocity set as an incoming stream with a periodic time change:

$$u_y(x) = 1.5[1+\beta\sin(2\pi f_i t)]\bar{U}\left(1 - \frac{y^2}{H_1^2}\right),$$
(16)

where $\beta$ is the amplitude of the incoming wave, and $f_i$ is the frequency of the incoming wave.

We can use reduced velocity to characterize the frequency of flow velocity change:

$$U_{red} = \bar{U}/f_i D. \tag{17}$$

When the average velocity of incoming flow and the diameter of the cylinder are constant, the reduced velocity is inversely proportional to the fluctuation frequency of incoming flow velocity.

**2.4 Numerical Validation (Vortex-induced vibration)**

There are less literature to simulate the flow induced vibration of a cylinder in expansion tube. We consider a uniform flow past a vibrating cylinder. The cylinder motion controlled by the following equation:

$$m\frac{d^2 x_i}{dt^2} + c\frac{dx_i}{dt} + kx_i = F_i(t), \tag{18}$$

where $x_i$ is the transversal displacement, $t$ is time, $F_i(t)$ is the transversal force acting on the cylinder, $m$, $c$ and $k$ represent the mass of the cylinder, damping and spring constants respectively. As the displacement and time are normalized by $D$ and $D/U_\infty$ respectively, the structural motion equation can be written as follow:

$$\frac{d^2 y^*}{dt^{*2}} + \frac{4\pi\xi}{U_{red}}\frac{dy^*}{dt^*} + \frac{4\pi^2}{U_{red}^2}y^* = \frac{2C_l}{\pi m^*}, \tag{19}$$

where $y^*$ and $t^*$ are the corresponding dimensionless transversal displacement and time. $U_{red}$ is the so-called reduced velocity which measures the root of the ratio of inertial force to spring force of the cylinder. It is defined by $U_{red} = U_\infty / (f_n D)$, where $f_n = 1/2\pi\sqrt{k/m}$ is the natural frequency of the cylinder. $m^* = 4\pi\rho D^2$ is the mass ratio of solid to fluid in the same volume, $C_l(t) = 2F_y(t)/(\rho l U_\infty^2 D)$ and $\xi = \frac{c}{2\sqrt{k/m}}$ are the corresponding lift and structural damping coefficients, respectively. The Reynolds number ($Re$) is set as 150. The reduced velocity ($U_{red}$) is from 3 to 12. The structural damping ratio ($\xi$) is set to zero. The Mass ratio ($M^*$) is set to 2.0.

The variation curve of the dimensionless maximum amplitude $A_y^*$ with $U_{red}$ is shown in Fig. 5. Our maximum $A_y^*$ is about 0.578$D$, which is close to the simulation performed by Xiong et al. (2019), Ahn &

Kallinderis (2006) and Borazjani & Sotiropoulos (2009). The phase diagram of $A_y^* \sim C_l$ for different reduced velocity ($U_{red}$) are shown in Fig. 5. Although there are some glitches on the curve of high $U_{red}$. The phase of $A_y^*$ and $C_l$ changes from a positive phase at low $U_{red}$ (such as $U_{red}$ =3.0) to an opposite phase at high $U_{red}$ (such as $U_{red}$ = 8.0). Our simulation can still capture this feature.

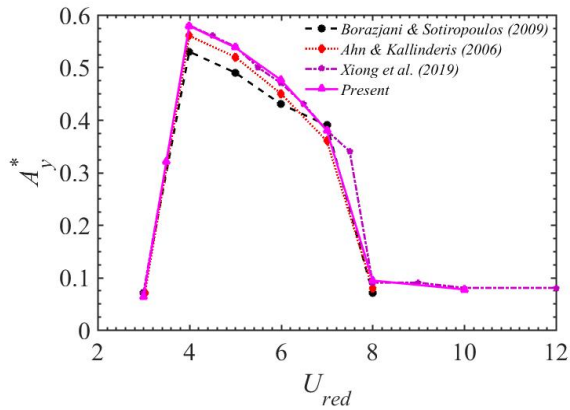

**Figure 5.** The amplitude curve of $A_y^* \sim U_{red}$.

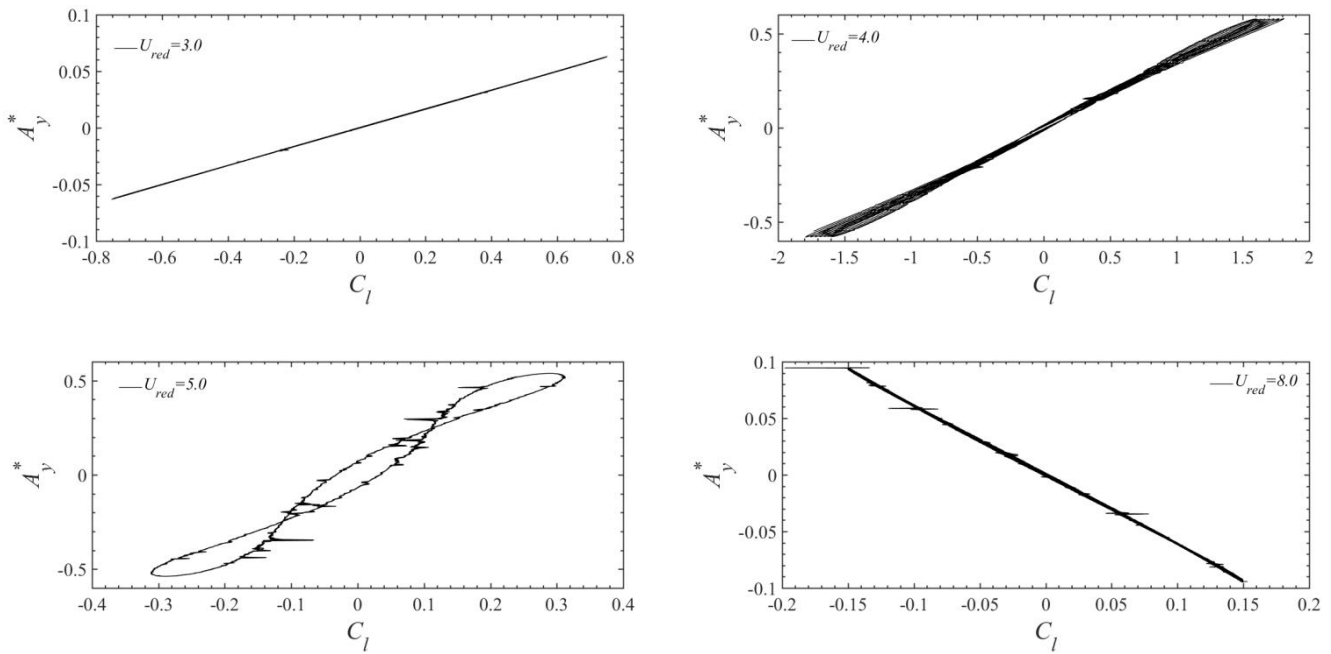

**Figure 6.** The phase diagram of $A_y^* \sim C_l$ for vortex-induced vibration.

## 3. Results and discussion

**3.1 invariable incoming flow**

Initially, we examine a invariable incoming flow, where we concentrate on two critical parameters:, dimensionless gravity ($Mg^*$) and Reynolds number ($Re$),.The values of $Mg^*$ and $Re$ are considered from 4.9 to 79.4 and 1 to 300, respectively. Tab. 3 lists the main motion form in these parameter space. Three movement patterns are found at the phase diagrams of ($Mg^*$, $Re$). For time-invariant incoming flow velocity profile, the phase diagrams can be divided into four regimes based on the movement of the cylinder. We will discuss them in detail below.

**Table 3**. The main motion form in our simulations

| $Mg^*\backslash Re$ | 1 | 2 | 3 | 4 | 5 | 10 | 20 | 40 | 100 | 200 | 300 |
|---|---|---|---|---|---|---|---|---|---|---|---|
| 4.9 | o | o | b | b | b | v | v | v | v | v | b |
| 9.8 | b | b | b | b | v | v | v | v | v | v | b |
| 19.6 | b | b | b | v | v | v | v | v | v | v | v |
| 39.25 | b | b | b | v | v | v | v | v | v | b | b |
| 78.4 | b | b | b | v | v | v | v | v | v | b | b |

**Note:** "o" represents flow out, "b" represents balance, "v" represents vibration, respectively.

For the flow around a cylinder, the thrust force ($F_y$) is negatively related with $Re$. For example, for an invariable uniform inflow flowing pass a confined cylinder in a free domain, the thrust force coefficient ($C_y$) decreases slowly with the increase of $Re$ from ~7 to ~1.2 in the range of $Re$ from 2 to 200 (Park et al., 1998; Qu et al., 2013; Li et al., 2022). At low $Re$, the thrust force may be larger than the cylinder's gravity. For example, at $Re = 1$, $Mg^* = 4.9$, the cylinder may go out at the dimensionless time $t^*$ is about 10, as shown in Fig. 7. The component mesh and background mesh are also displayed.

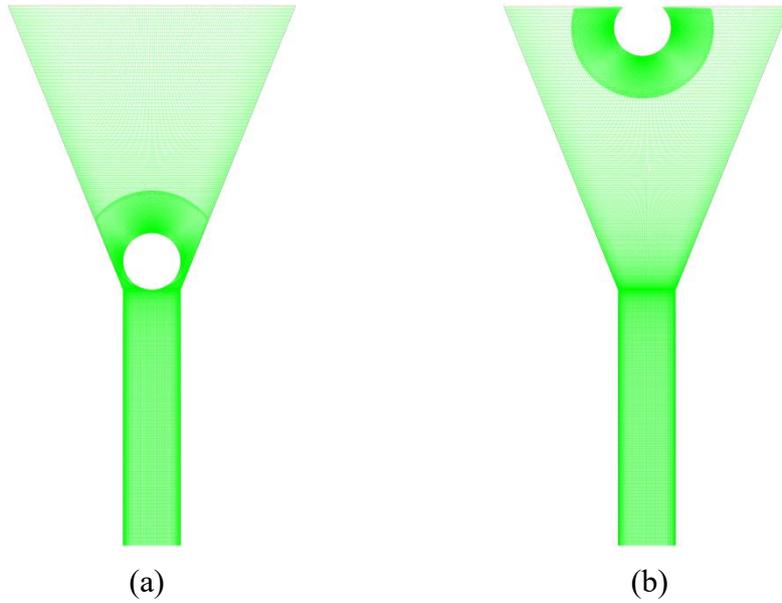

(a)                              (b)

**Figure 7.** The cylinder's component mesh and background mesh for (a) $t^* = 0$, (b) $t^* = 10$ for $Re = 1$, $Mg^* = 4.9$.

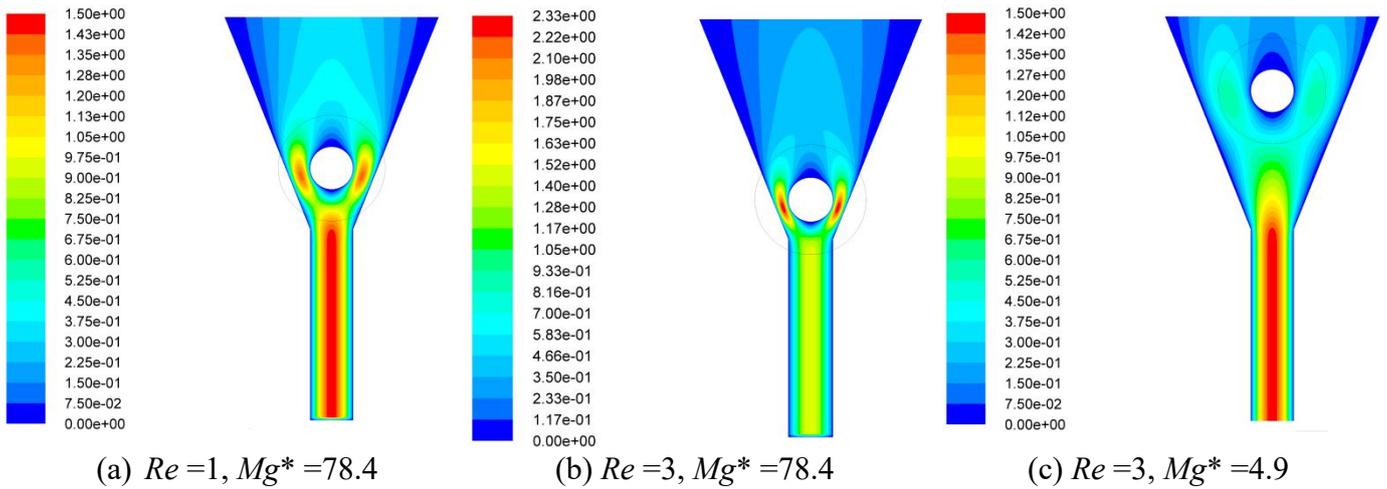

(a) $Re = 1$, $Mg^* = 78.4$        (b) $Re = 3$, $Mg^* = 78.4$        (c) $Re = 3$, $Mg^* = 4.9$

**Figure 8.** Velocity distribution in equilibrium.

As the Reynolds number increases or the dimensionless gravity ($Mg^*$), the thrust force acting on the cylinder may eventually balance the cylinder's gravity. At specific combinations such as $Re = 1$, $Mg^* = 79.4$, $Re = 3$, $Mg^* = 79.4$ and $Re = 3$, $Mg^* = 4.9$, the cylinder finally be fixed a point. We show the velocity profile of these three cases at the balance point in Fig. 8. When $Mg^*$ is fixed at 78.4, increasing $Re$ may cause the cylinder to move towards the bottom of the expansion tube, resulting in a reduction in the flow's thrust. Conversely, for a constant $Re$, decreasing $Mg^*$ could lead to the cylinder moving upward.

In Tab. 3, there exists an interval, approximately between $Re$ from 4 to 200, where the cylinder may exhibit vibrations. The trajectory of the vibration resembles the shape of the numeral '8' within the tube. As the cylinder approaches the bottom of the funnel, the amplitude of the cylinder's vibration becomes smaller. This phase often corresponds to larger Reynolds number and mass ratio. The Reynolds number of vortex shedding appears around 47.5 when a fixed cylinder flows in a free domain. However, in cases where the cylinder vibrates in the tube, the onset of vortex shedding appears much earlier. At high Reynolds numbers, the flow tends to become more turbulent, though this behavior is not consistent, and no clear periodic orbits are evident in the trajectory curve. In some instances, the flow unexpectedly stabilizes, highlighting the need for further investigation.

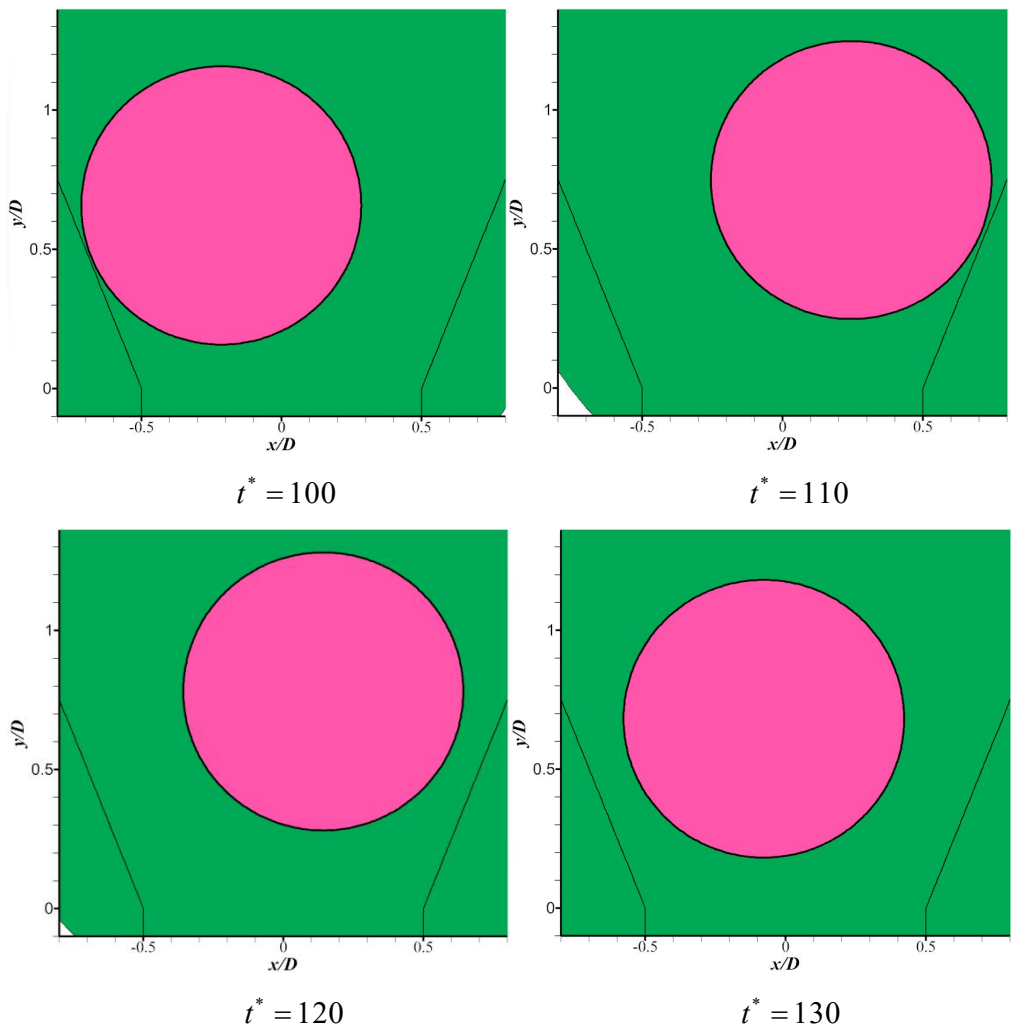

**Figure 9.** The motion for *Re* =10, *Mg\**=19.6.

At high Reynolds numbers, distinct equilibrium states are observed with low and high weights. At lower weight, the cylinder will balance on the central axis. Conversely, at high weights, the cylinder is not balanced on the central axis. And asymmetric jets appear on both sides of the cylinder. It is possible that the central axis is not a stable equilibrium point towards the bottom of the cylinder, warranting further investigation.

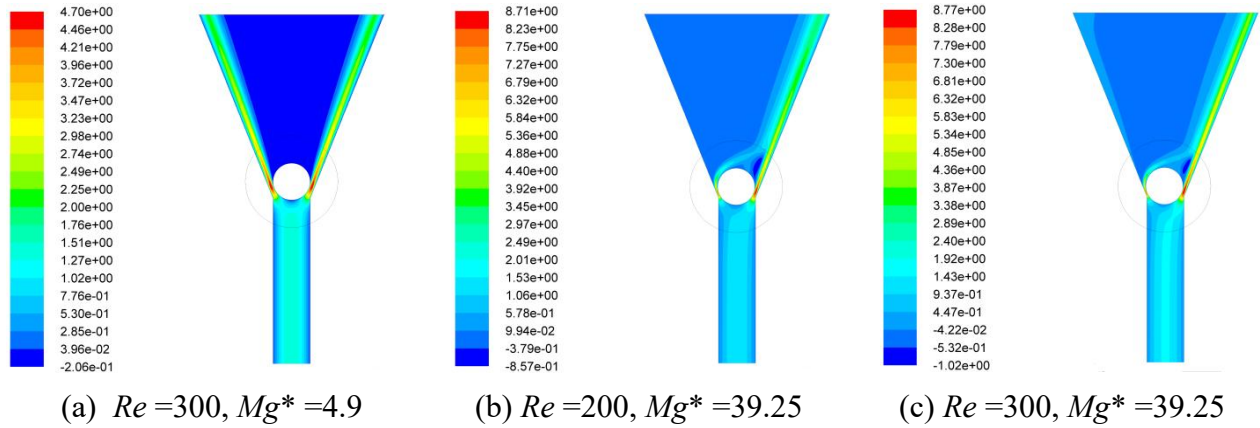

(a) $Re$ =300, $Mg^*$ =4.9    (b) $Re$ =200, $Mg^*$ =39.25    (c) $Re$ =300, $Mg^*$ =39.25

**Figure 10.** Balance at high Reynolds numbers.

The time-averaged *y*-coordinates of the cylinder centers are listed in Tab. 4. Obviously, the time-averaged *y*-coordinate are negatively related to Reynolds number ($Re$) and dimensionless gravity ($Mg^*$).

**Table 4**. Y-coordinate of the average balance point.

| $Mg^*\backslash Re$ | 1 | 2 | 3 | 4 | 5 | 10 | 20 | 40 | 100 | 200 | 300 |
|---|---|---|---|---|---|---|---|---|---|---|---|
| 4.9 | - | - | 3.254 | 2.804 | 2.497 | 1.763 | 1.051 | 0.934 | 0.700 | 0.615 | 0.582 |
| 9.8 | 3.943 | 2.810 | 2.285 | 1.980 | 1.561 | 1.118 | 0.805 | 0.642 | 0.539 | 0.479 | 0.455 |
| 19.6 | 2.816 | 1.988 | 1.638 | 1.423 | 1.145 | 0.860 | 0.630 | 0.503 | 0.423 | 0.381 | 0.362 |
| 39.25 | 1.992 | 1.439 | 1.201 | 0.992 | 0.855 | 0.644 | 0.487 | 0.398 | 0.337 | 0.304 | 0.291 |
| 78.4 | 1.443 | 1.065 | 0.899 | 0.751 | 0.665 | 0.495 | 0.395 | 0.301 | 0.272 | 0.246 | 0.236 |

The Study on Movement of Cylinder in Expansion Tube

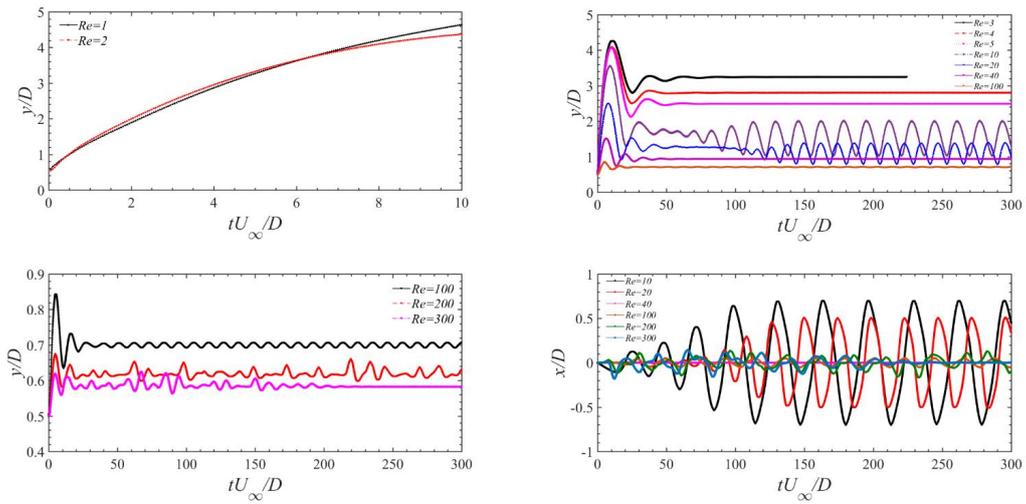

(a) $Mg^* = 4.9$

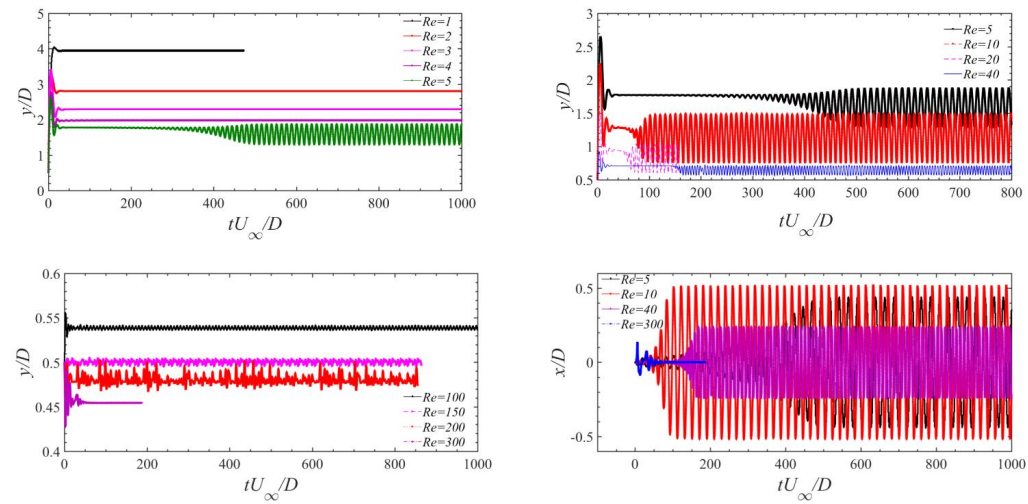

(b) $Mg^* = 9.8$

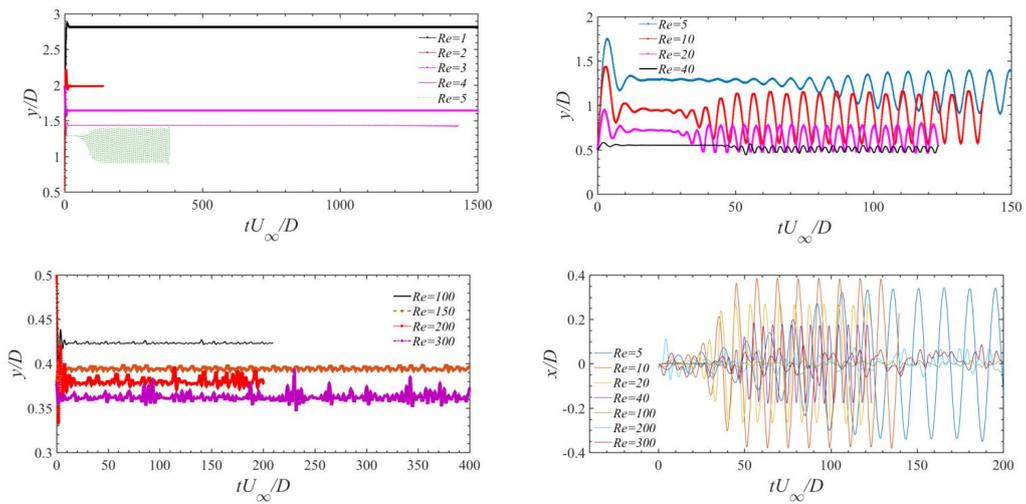

(c) $Mg^* = 19.6$

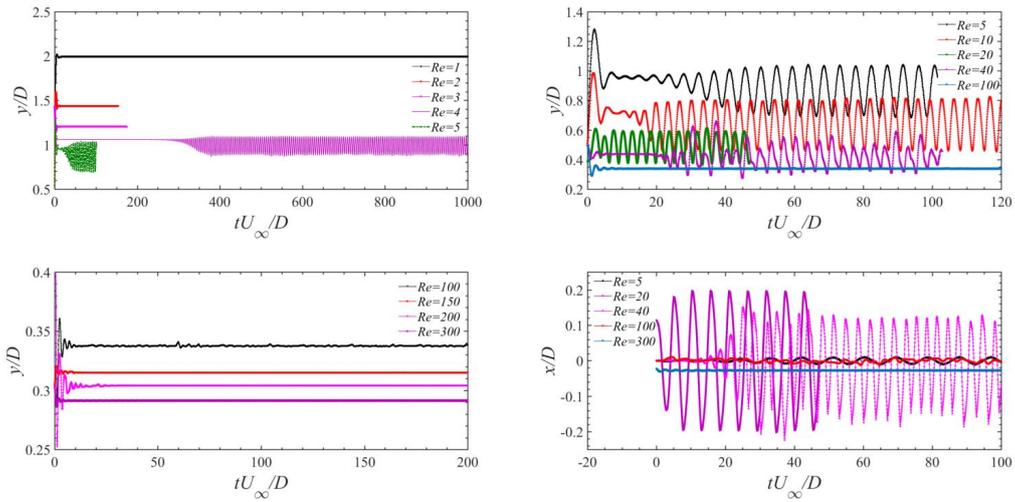

(d) $Mg^* = 39.25$

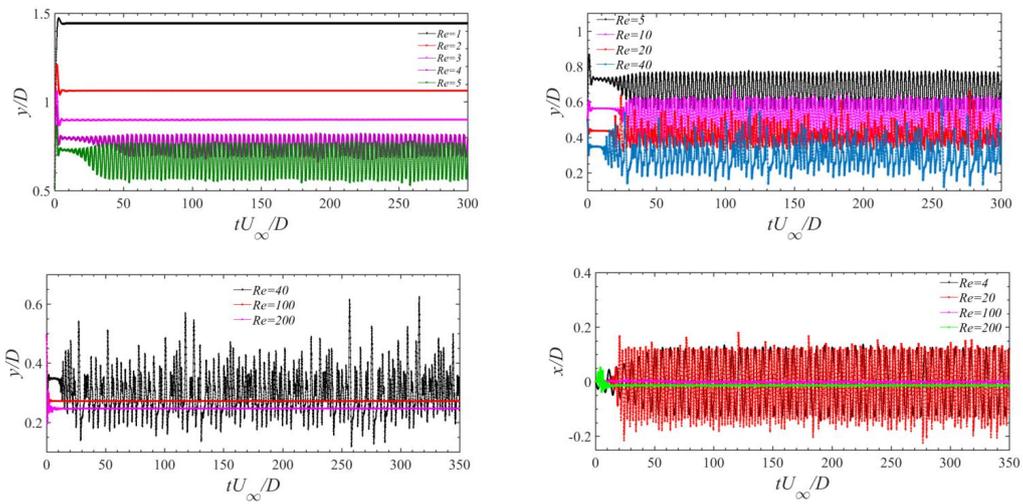

(e) $Mg^* = 78.4$

**Figure 11.** Movements over time.

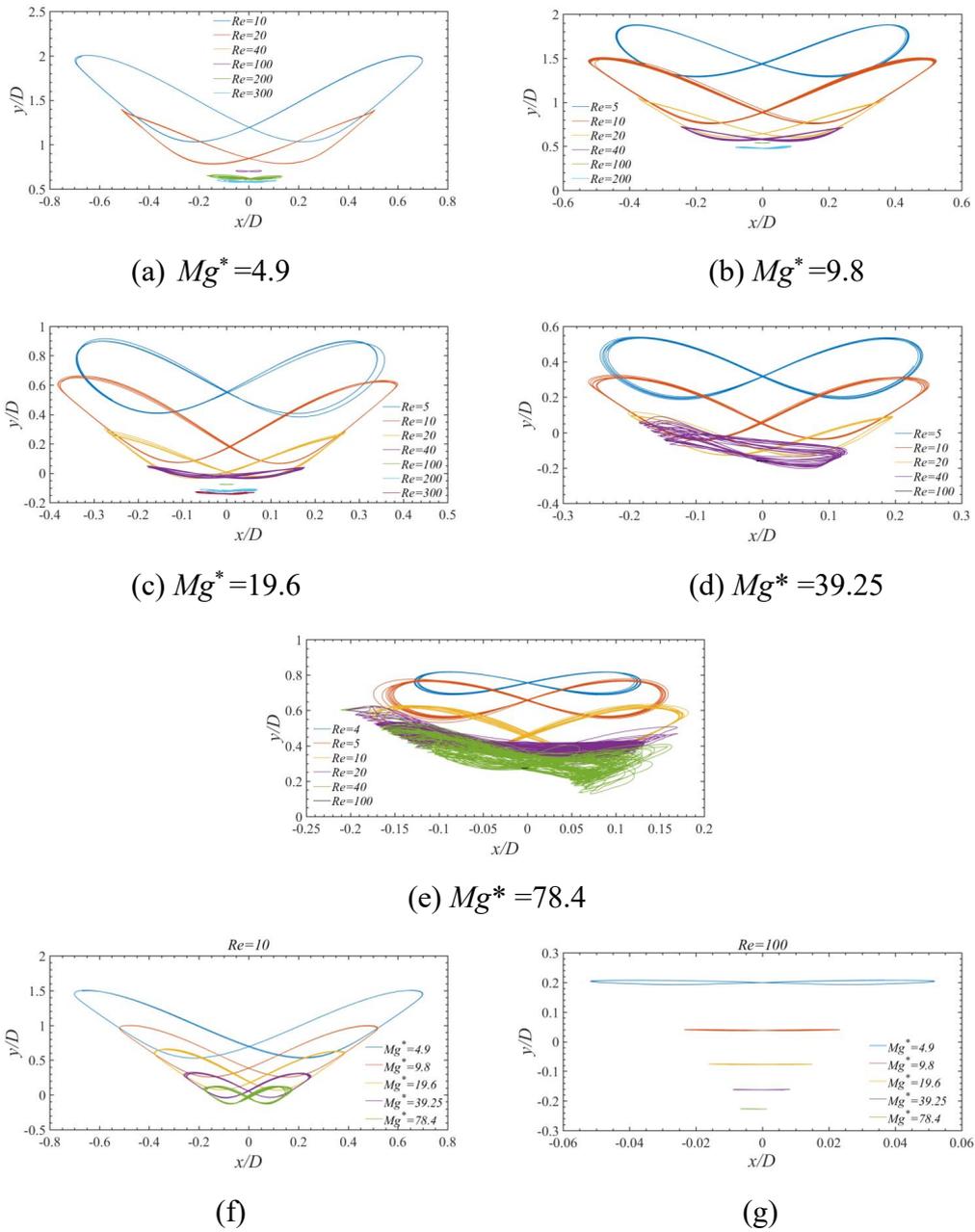

**Figure 12.** Trajectory of $y/D \sim x/D$.

As the Reynolds number continues to increase, the uniform balance point of the cylinder moves to the bottom of the expansion tube, and then the increase gradually weakens, which may be due to the wall restriction of the expansion tube. However, at $Re=200$ and $Re=300$, the vibration pattern no longer displays pronounced sinusoidal periodic motion. Other cases with varied $Mg^*$ values behave similarly to $Mg^*=4.9$. with larger $Mg^*$, preventing the cylinder from exiting even at $Re=1$. The critical Reynolds number ($Re_c$) of the initial vibration of the cylinder is found to be smaller as $Mg^*$ becomes larger. Such, $Re_c$ is between 5 to 10 for $Mg^*$ =4.9. Relatively, it is between 3 to 4 for $Mg^*$ =78.4. For all $Mg^*$, the maximum amplitude occurs at approximately Reynolds number 10. At high Reynolds numbers, the vibration all tends to become weak. The trajectory of $y/D \sim x/D$ for different cases are depicted in Fig. 12. The majority of trajectories exhibit an "8" shape, with the asymptotic line being the expansion tube. But there are some exceptions, such as $Mg^*$ =39.25, $Re$ =40 and $Mg^*$ =78.4, $Re$ =40, the trajectory of the vibration are no longer like the "8" shape.Instead, they deviate from periodicity and shift towards chaotic trajectories.

**3.2 Sinusoidal periodic fluctuation incoming flow**

We select the case of $Re$ =300, $Mg^*$ =39.25 to superimpose a sinusoidal periodic fluctuation in the incoming flow expressed by Eqn. (16). Although under these parameters it had been previously determined that the cylinder would come to a stop off the central axis in a uniform stream, our latest findings, as depicted in Fig. 13,    suggest that the cylinder may exhibit continuous vibration with a complex trajectory when subjected to our specified alternating inlet flow. Notably, the ordinate of the center point of the trajectory almost all at $\overline{y}/D$ =0.291, which seems only related to the time average incoming stream. Two key parameters effects the non-uniform incoming stream: $\beta$ and $U_{red}$, affecting the fluctuation amplitude and frequency, respectively. A larger $\beta$ corresponds to the larger the amplitude of fluctuation, while a higher $U_{red}$ means a longer fluctuation period. For $\beta$ =0.5, $\beta$ =0.9 and 0.99, Reynolds number ($Re$) range from 150 to 450, 27 to 570 and 3 to 597 for transient flow. For $\beta$ =0.5, $U_{red}$ =8.0, the mid-line point of trajectory deviates significantly from the position of $x$ =0, which seems to the case of uniform incoming stream. In this case, the cylinder does not come to a complete standstill but rather exhibits periodic vibration within the expansion tube. For other cases, the mid-line point of trajectory all at $x$ =0. For higher $U_{red}$ or $\beta$, the cylinder tends to have a larger vibration amplitude in the $y$ direction. The cylinder is more likely to appear at the bottom of the expansion tube.

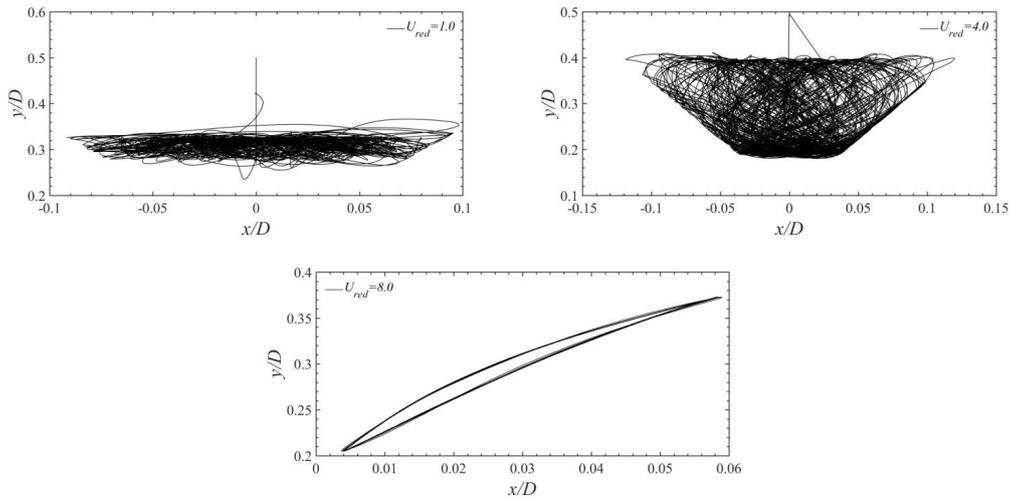

(a) $Re = 300$, $Mg^* = 39.25$, $\beta = 0.5$

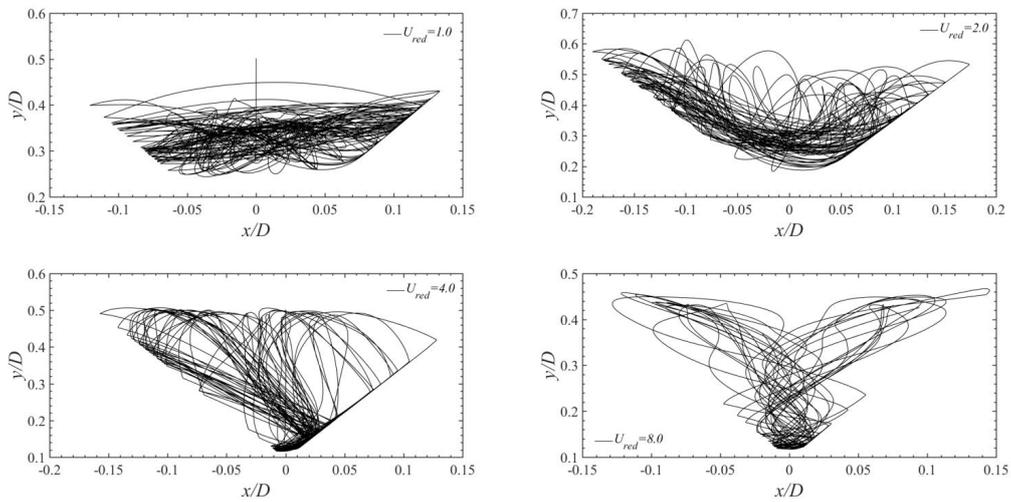

(b) $Re = 300$, $Mg^* = 39.25$, $\beta = 0.9$

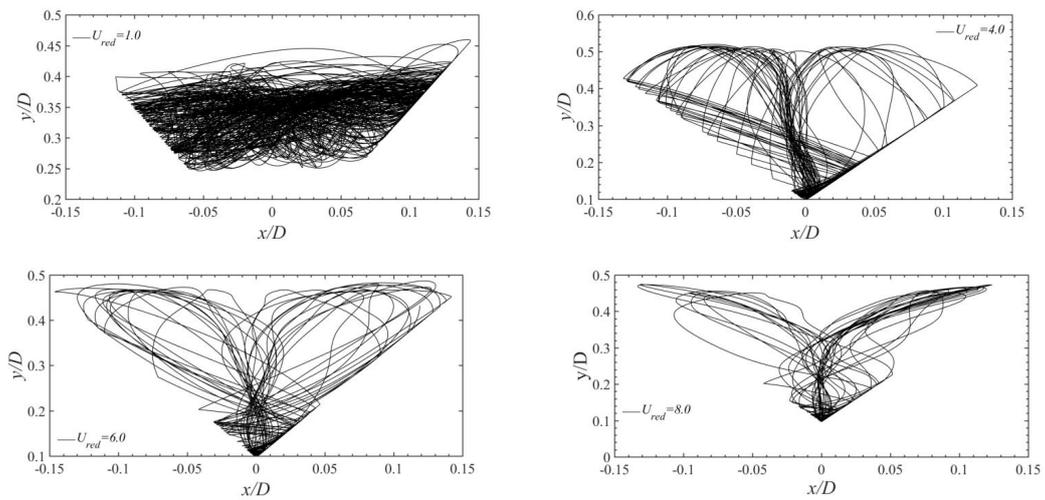

(c) $Re = 300$, $Mg^* = 39.25$, $\beta = 0.99$
**Figure 13.** The trajectory of volatility inlet velocity.

Additionally, for higher $\beta$, the incoming stream have wide rage $Re$, resulting in a wider range balance point. Meanwhile, higher $U_{red}$, correspond to longer fluctuation periods, consequently extending the duration of the same Reynolds number interval and leading to larger vibration amplitudes.

## 4. Conclusion

In this study, series two-dimensional simulations are obtained to understand the movement of the small cylinder in expansion tube. These simulations are performed at FLUENT platform ,incorporating the Overset function. The collision between the cylinder and the tube's wall are considered as a positive collision of rigid body without energy loss. Two key parameters, dimensionless gravity ($Mg^*$) and Reynolds number ($Re$), are focused on. $Mg^*$ and $Re$ are considered from 4.9 to 79.4 and 1 to 300, respectively. Two types of incoming flow are considered: uniform incoming flow and sinusoidal periodic fluctuation incoming flow. Three movement patterns are found at the phase diagrams of ($Mg^*$, $Re$). For uniform incoming flow, the phase diagrams can be divided into four regimes based on the movement of the cylinder. At low $Mg^*$ and low $Re$, such as $Mg^*$ =4.9, $Re$ =1, the cylinder may flow out for high thrust. Conversely, at higher $Mg^*$ and $Re$, such as $Mg^*$ =4.9, $Re$ =4 and $Mg^*$ =19.6, $Re$ =3, the cylinder may reach a stable position through a balance of gravity and thrust. In scenarios with higher $Mg^*$ and $Re$, such as $Mg^*$ =9.8, $Re$ =5, the cylinder may be vibration in the expansion tube for the cylinder's wake shedding. Notably, at high $Mg^*$ and $Re$, such as $Mg^*$ =78.4, $Re$ =300, a strange phenomenon is discovered, the small ball may not eventually be vibration instead of still at a point where deviate from the center line. High-amplitude vibration of cylinder are prevalent in the range of Reynolds number from 5 to 40 for the expansion tube flow of uniform incoming flow. Furthermore, the study also considered $Re$ of 300 with sinusoidal periodic fluctuation incoming flow, demonstrating that, unlike uniform flow, the cylinder vibrated significantly under this condition. It is important to note that the two-dimensional cylinder model simplifies the behavior of a sphere when

00

disregarding its three-dimensional motion. Overall, this research offers valuable insights into the underlying mechanisms governing the movement of the cylinder under various flow conditions.

## 5. Acknowledgments

This research was funded by the Applied Basic Research Project of Sichuan Province (2022NSFSC1988). This work is supported by Center for Computational Science and Engineering of Southern University of Science and Technology.